# A Multi-Trait Approach Identified Genetic Variants Including a Rare Mutation in RGS3 with Impact on Abnormalities of Cardiac Structure/Function

Akram Yazdani[1*], Azam Yazdani[2], Raúl Méndez Giráldez[3], David Aguilar[4], Luca Sartore[5]

[1]Department of Genetics and Genomic Science, Icahn School of Medicine at Mount Sinai, New York, NY, USA
[2]Section of Preventive Medicine and Epidemiology, School of Medicine, Boston University, MA, USA
[3]Lineberger Comprehensive Cancer Center, University of North Carolina School of Medicine, Chapel Hill, NC, USA
[4]Baylor College of medicine, Houston, TX, USA
[5]National Institute of Statistical Science/National Agriculture Statistics Service, Washington, DC, USA

*To whom correspondence should be addressed.
Email address: akram.yazdani@mssm.edu, akramyazdani16@gmail.com

## Abstract

Heart failure is a major cause for premature death. Given heterogeneity of the heart failure syndrome, identifying genetic determinants of cardiac function and structure may provide greater insights into heart failure. Despite progress in understanding the genetic basis of heart failure through genome wide association studies, heritability of heart failure is not well understood. Gaining further insights into mechanisms that contribute to heart failure requires systematic approaches that go beyond single trait analysis.

We integrated Bayesian multi-trait approach and Bayesian networks for the analysis of 10 correlated traits of cardiac structure and function measured for 3387 individuals with whole exome sequence data. While using single-trait based approaches did not find any significant genetic variant, applying the integrative Bayesian multi-trait approach, we identified 3 novel variants located in genes, *RGS3*, *CHD3*, and *MRPL38* with significant impact on the cardiac traits such as left ventricular volume index, parasternal long axis interventricular septum thickness, and mean left ventricular wall thickness. Among these, the rare variant NC_000009.11:g.116346115C>A (rs144636307) in *RGS3* showed pleiotropic effect on left ventricular mass index, left ventricular volume index and Maximum left atrial anterior-posterior diameter while *RGS3* can inhibit TGF-beta signaling associated with left ventricle dilation and systolic dysfunction.

## Introduction

Heart failure (HF) is a complex clinical syndrome characterized by abnormal cardiac structure and function that leads to reduced cardiac output and elevated filling pressures at rest or with exertion (1). Although, there is increasing evidence that the risk and course of HF depend on genetic predispositions (2), genome wide association studies (GWAS) have identified only a handful of genetic variants associated with it. For instance, the chromosome region 9p21 includes several highly replicated genetic variants associated with HF risk factors (e.g. NC_000009.11:g.22096055A>G and NC_000009.11:g.22124477A>G) (3). The effect allele G of the first variant is intronic to gene *CDKN2B-AS1* while the same effect allele in the second variant is located in an enhancer and disrupts a binding site for *STAT1* (4).

For better understanding heritability of HF, some studies combine results of multiple cohort and involve more samples in the analysis through meta-analysis (5). One of the largest studies of these on African-American population identified four variants including NC_000008.10:g.49171227A>G (rs4552931) and NC_000017.10:g.66685847C>T (rs7213314) associated with left ventricular mass and left ventricular internal diastolic diameter respectively using Echocardiography (6). A meta-analysis on 5 cohorts of individuals with European ancestry identified five genetic loci harboring common variants associated with left ventricular diastolic dimensions and aortic root size (7). More recently, a meta-analysis through studying a large set of samples including 73,518 individuals identified 32 novel loci associated with electrocardiographic markers of hypertrophy as an important and independent risk factor for the development of heart failure (8).

Taking advantage of dozens-to-hundreds of traits measured on each study participant creates opportunities to obtain insights into systems biology of HF, and consequently reduces morbidity, and economic burden of HF. Multi-trait analysis is toward this aim and increases the statistical power (9)(10)(11)(12). Although there are many studies on multi-trait approaches, applications of those methods have recently received increased attention e.g. (13)(14)(15). A limitation of those methods is their complexity due to the large number of parameters in the model.

To reduce the complexity of the multi-trait models due to the number of parameters, we here integrated Bayesian network (16) with a Bayesian polygenic mixed model while setting G-Wishart prior on covariance matrix (17) and called it Integrative Bayesian Multi-Trait (IBMT) approach. Using IBMT approach, we conducted an analysis to identify genomic variants influencing 10 echocardiographic traits related to cardiac structure and function from Atherosclerosis Risk in Communities (ARIC) study (18). We have genotype data of 7810 European American individuals from baseline measurement while 3387 of these individuals have phenotype records in Visit 5. The phenotype data are also recorded for 1265 new participants at visit 5 who do not have baseline genotype. Thus, we had three sets of data, including i) individuals with only genotype data, ii) individuals with only phenotype data, and iii) individuals with both genotype and phenotype data. We incorporated information from all these three sets into the analysis to improve the statistical power, prevent overfitting, and avoid using data multiple times. The details are provided in the Methods section. These steps ultimately improve reliability and generalizability of the results.

After data preparation, we applied the IBMT method over the whole exome sequence data to investigate genomic and cardiac trait relationships. We identified 3 genetic variants (NC_000009.11:g.116346115C>A, NC_000017.10:g.7802658C>T, NC_000017.10: g.73897977 C>T) in genes *RGS3*, *CHD3*, and *MRPL38* with significant impact on the cardiac traits such as left ventricular volume index, parasternal long axis interventricular septum thickness, and mean left ventricular wall thickness. Gene *RGS3* codes for GTPase-activating protein that inhibits G-protein-mediated signal transduction. *CHD3* encodes a protein with a chromatin organization modifier domain and a SNF2-related helicase/ATPase domain. *MRPL38* produces a mitochondrial ribosomal protein, involved in the synthesis of proteins within the mitochondrion.

The variant in *RGS3* gene (NC_000009.11:g.116346115C>A) showed pleiotropic gene action on vertical mass index, left ventricular volume index and Maximum left atrial anterior-posterior diameter) while *RGS3* can inhibit TGF-beta signaling associated with left ventricle dilation and systolic dysfunction.

## Materials and Methods

### Study population

Echocardiographic and genomic data were collected on a subset of the ARIC study, a biracial longitudinal cohort of 15,792 middle-aged individuals who were randomly sampled from four US sites (Forsyth County, NC; Jackson, MS; suburbs of Minneapolis, MN; and Washington County, MD) and have been measured for risk factor traits related to health and chronic diseases. A detailed description of the ARIC study design and methods have been published elsewhere (18)(19). The data presented here includes 7810 European American individuals with baseline genotype available on dbGAP (https://www.ncbi.nlm.nih.gov/gap), accession number phs000090.v1.p1. A subset of individuals with genotype data, 3387 out of 7810 individuals, has phenotype records at visit 5. The phenotype data described in the following subsection are also recorded for 1265 new participants at visit 5 who do not have baseline genotype. Figure 1 visualized study population with phenotype and genotype records through Venn diagram.

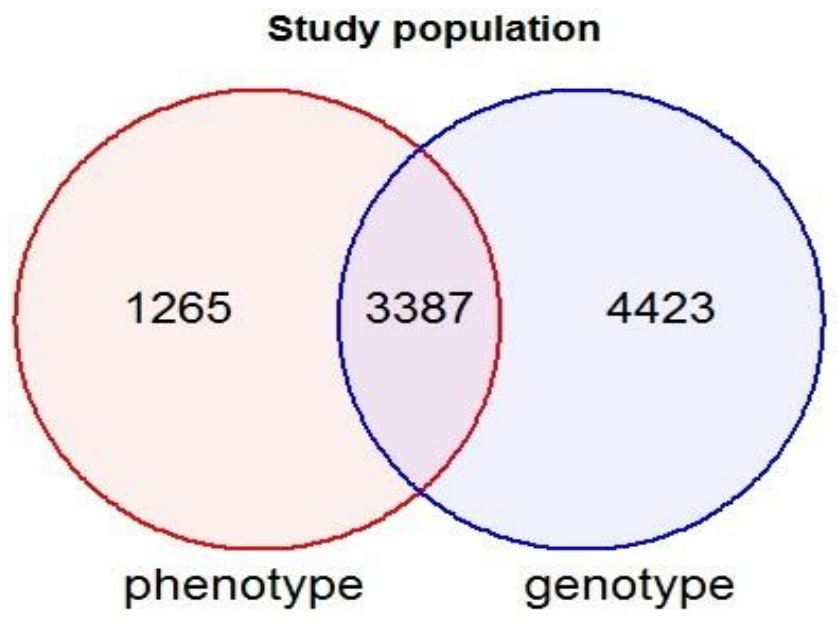

**Figure 1.** Venn diagram of study population with phenotype and genotype records.

## Echocardiographic methods and measurements

Echocardiograms were obtained from participants at visit 5 using a standardized protocol as recommended by the American Society of Echocardiography. Images were digitally transferred to the Cardiovascular Imaging Core Laboratory at Brigham and Women's Hospital, Boston, MA, for offline analysis. The intra-observer variability (coefficient of variation and interclass correlation) for key echocardiographic measures has been previously published (20). Images were obtained in the parasternal long- and short-axis and apical 2- and 4-chamber views. Primary measures of the traits such as left ventricular (LV) dimensions, volumes, and wall thickness; left atrial (LA) dimension, volume, and area were made in triplicate from the 2-dimensional views in accordance with the recommendations of the American Society of Echocardiography (21). This study includes 10 cardiac structure and function tabulated in Table 1. LV mass was calculated from LV linear dimension and indexed to body surface area. Relative wall thickness was calculated using the posterior wall thickness and LV end-diastolic dimension. LA volumes were measured by methods of disks using apical 4- and 3- chamber views. LV volumes were calculated from the apical 4- and 2- chamber views utilized the modified Simpson method.

**Table 1.** Cardiac structural and functional traits under study.

| Phenotypes of Interest | |
|---|---|
| **Name** | **Measurement** |
| Parasternal long axis interventricular septum thickness (PLAx-IST) | Cm |
| Parasternal long axis posterior wall thickness (PLAx-PWT) | Cm |
| End-diastolic volume (ED-V) | Ml |
| End-systolic volume (ES-V) | Ml |
| Ejection fraction (EF) | % |
| LV mass index (LV-MI) | G per m2 |
| LV relative wall thickness (LV-RWT) | |
| Mean LV wall thickness (LV-WT) | Cm |
| Maximal left atrial anterior-posterior diameter (Max-LA-APD) | Cm |
| LA volume index (LA-VI) | Ml per m2 |

## Whole exome sequencing

Whole exome sequencing was performed on samples with the Illumina HiSeq platform. Mercury pipeline is applied for variant calling (22). The reads are mapped into the Genome Reference Consortium Human Build 37 (GRCh37) sequence. Low-quality variants are filtered if they were outside the exon capture regions, belonged to multi-allelic sites, had missing rate > 20% and had mean depth of coverage > 500-fold. In addition, highly significant departures from race-specific Hardy-Weinberg equilibrium (P-value < 5e-6) are excluded from the data.

## Statistical methods

### Adjusting covariates

There is a broad consensus on analytic techniques for covariate adjustment to discover genomic variants associated with traits of interest independently of the correlated covariates and improve statistical power by gaining precision. The set of covariates typically include principal components of individual genotypes to account for population structure, correlated environmental or demographic factors such as gender and age. However, some of those covariates may not have significant impact on the traits and the adjustment may contaminate the data. Therefore, instead of adjusting each trait for all routine covariates, we first investigated the effect of covariates on the traits using 1265 individuals with only phenotypic record. This not only prevents decreasing the statistical precision or contaminating data due to adjusting for unrelated covariates but also avoids using data twice which prevents overfitting.

### Applying an Integrative Bayesian Multi-Trait approach

We applied Empirical Bayesian approach for analysis of multiple correlated traits. We first identified underlying relationships of the 10 cardiac functional and structural traits tabulated in Table 1 via application of Bayesian networks. The analysis was carried out at a statistical significance level of 0.05 determined by structural Hamming distance (23)(24)(25). The identified network, which is also supported by clinical background knowledge, revealed the sparsity level of the relationships. Figure 2 displays the network, where the nodes are the traits and the edges represent a significant relationship between the two corresponding traits after excluding the effect of the other traits. Clustering approaches (26)(27) can be also applied for the same purpose, although they may estimate more connections among the traits. The Bayesian network was built on the subset of individuals with only phenotype records to avoid overfitting.

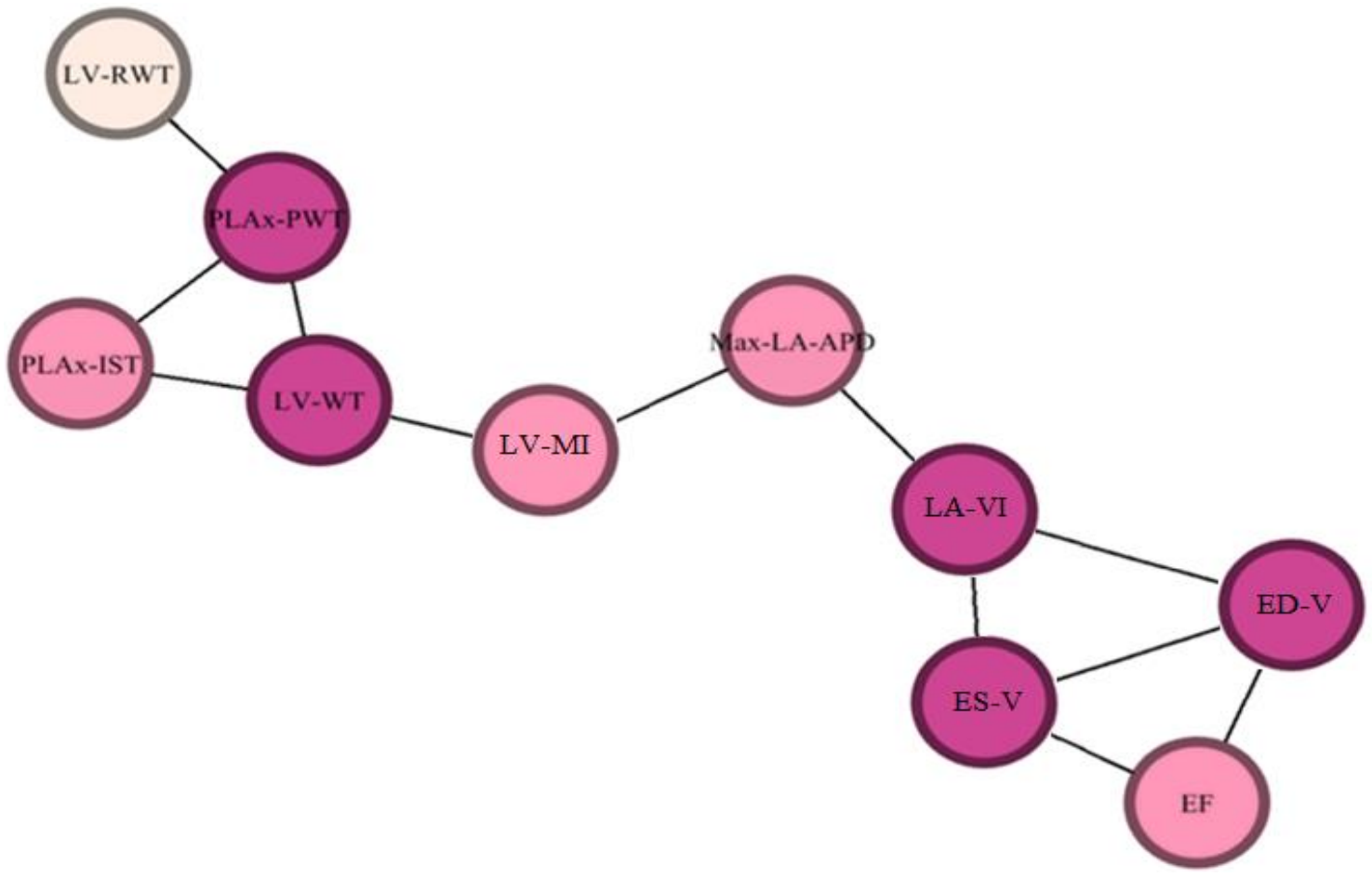

**Figure 2.** The Bayesian Network over the cardiac traits. The colors correspond to the degree of connectivity of each trait; deeper color means greater connectivity.

We integrated the identified underlying relationships among the traits with a Bayesian polygenic mixed model while setting G-Wishart prior on covariance matrix and called it Integrative Bayesian Multi-Trait (IBMT) approach. Applying the IBMT, we optimized performance and computational time of Gibbs Sampling Scheme due to the decrease on the number of parameters of the model. Further details of the model and Gibbs sampling scheme are provided in the Supplementary statistical methods section.

Using the identified Bayesian network over the traits, we implemented the IBMT method and analyzed the whole exome sequencing data including 260,688 variants based on sliding window. Each window included 100 variants with a step size of 25, such that each variant appears in 4 windows. If a variant is selected in all 4 windows, we reported the variant as the most promising variants. The parameters of the model were updated at each iteration of Markov Chain Monte Carlo (MCMC) algorithm and estimated with posterior means, which calculate optimal point estimators under square error loss, after 200 burn-in period.

To identify genomic variants significantly associated with the traits, we calculated 98% credible interval (28)(29), $(q^{\mathrm{L}}, q^{\mathrm{U}})$, for the effects of all the variants to test a null hypothesis of no genomic effects. The endpoints of the intervals correspond to quantile ($q$) of the empirical distribution of the MCMC drown from the marginal posterior distribution of genetic effects. A desired degree of precision for the endpoints of intervals is achieved by running a number of iterations until

$$P\left(\left|q_i^L - q_{i-1}^L\right|\right) < \zeta, \qquad P\left(\left|q_i^U - q_{i-1}^U\right|\right) < \zeta,$$

where $q^{\mathrm{L}}$ and $q^{\mathrm{U}}$ are lower and upper quantile respectively and $i$ represents the number of iterations. We set ζ to 0.01 as a small value.

**Results**

Since the IBMT method is based on a linear polygenic mixed model, we first tested the normality of the traits as a fundamental assumption. Except ejection fraction and maximal left atrial anterior-posterior diameter that are normally distributed, the other traits have been transformed to normal using log transformation. The histograms of the traits after winsorization, standardization, and log transformation are represented in Supplementary Figure1.

We then investigated the effects of gender, age, ever-smoked, body mass index (BMI), hypertension, systolic and diastolic blood pressure on the traits. Among them gender and BMI showed highly significant relationships (P-value < 1e-8) with all traits except mean LV Wall Thickness (LV-WT) and ejection fraction, which they were relatively less significant with, P-values 0.05 and 0.005 respectively. Hypertension also showed significant effects (P-value < 1e-6) with all traits except ejection fraction. We obtained these results on the set of individuals without genotype data to avoid the use of data twice. To generalize the results to the set of interest (individuals with both genotype and phenotype records), we compared BMI distribution, gender ratio (Female/Male), and ratio of (with/without) hypertension in the two sets. We observed that in both sets, the distribution of BMI is similar (Figure 3); gender ratios 1.302 and 1.387 showed almost the same proportion of female to male; and hypertension ratios 2.502 and 2.579 also showed almost the same proportion of individuals with and without hypertension. Therefore, we adjusted the traits for BMI, gender, and hypertension, in addition to the first 10 PCs from population stratification analysis.

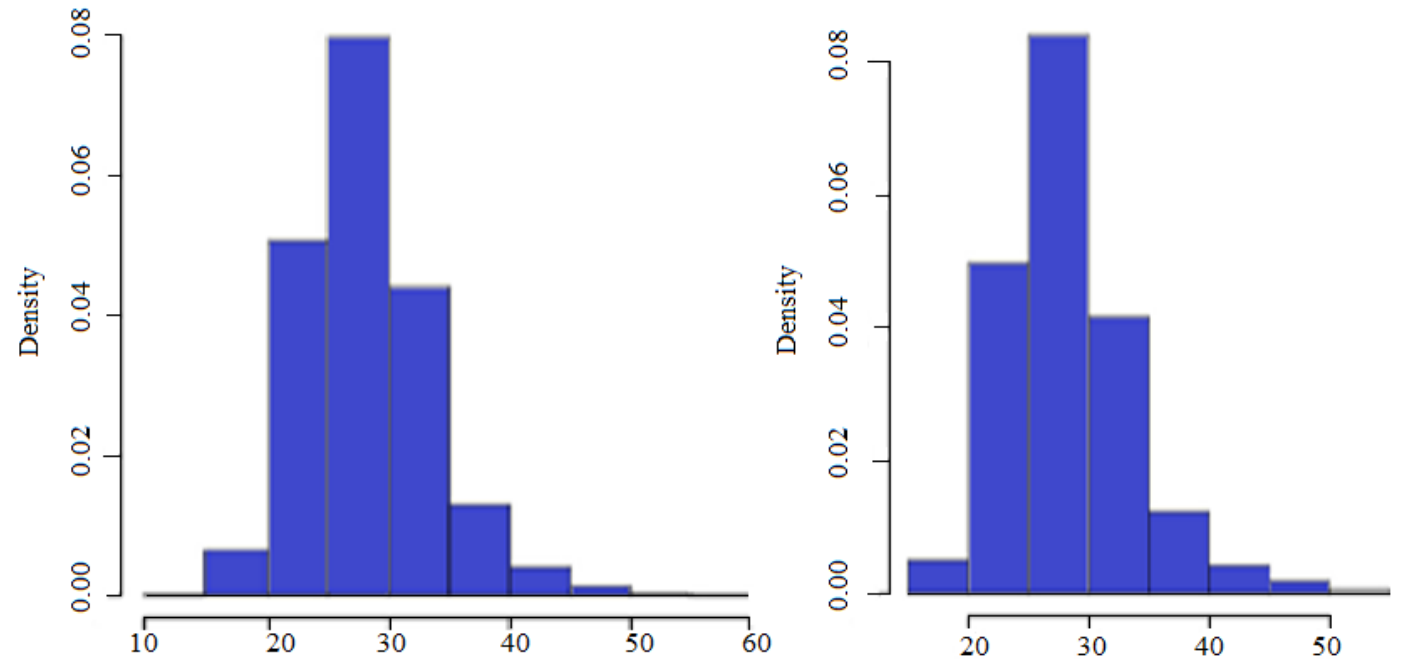

**Figure 3.** Histogram of BMI. Right: individuals without genotype data. Left: individuals with genotype data.

Applying IBTM, we first identified Bayesian network among 10 traits listed in Table 1. As shown in Figure 2, underlying relationships among the traits are sparse. Therefore, we do not need to consider all pairwise connectivity in the analysis. Incorporating this result into multi-trait mixed model, we reduced the number of parameters in the model and consequently increased the power of genotype-phenotype identification (Supplementary Statistical methods). We could identify 3 genetic variants with significant impact on 4 cardiac traits using a 98% credible interval (Table 2 and Figure 4). Minor allele frequencies (MAF) in Table 2 show the identified variants are rare. Table 3 reports the estimated effects of the identified variants.

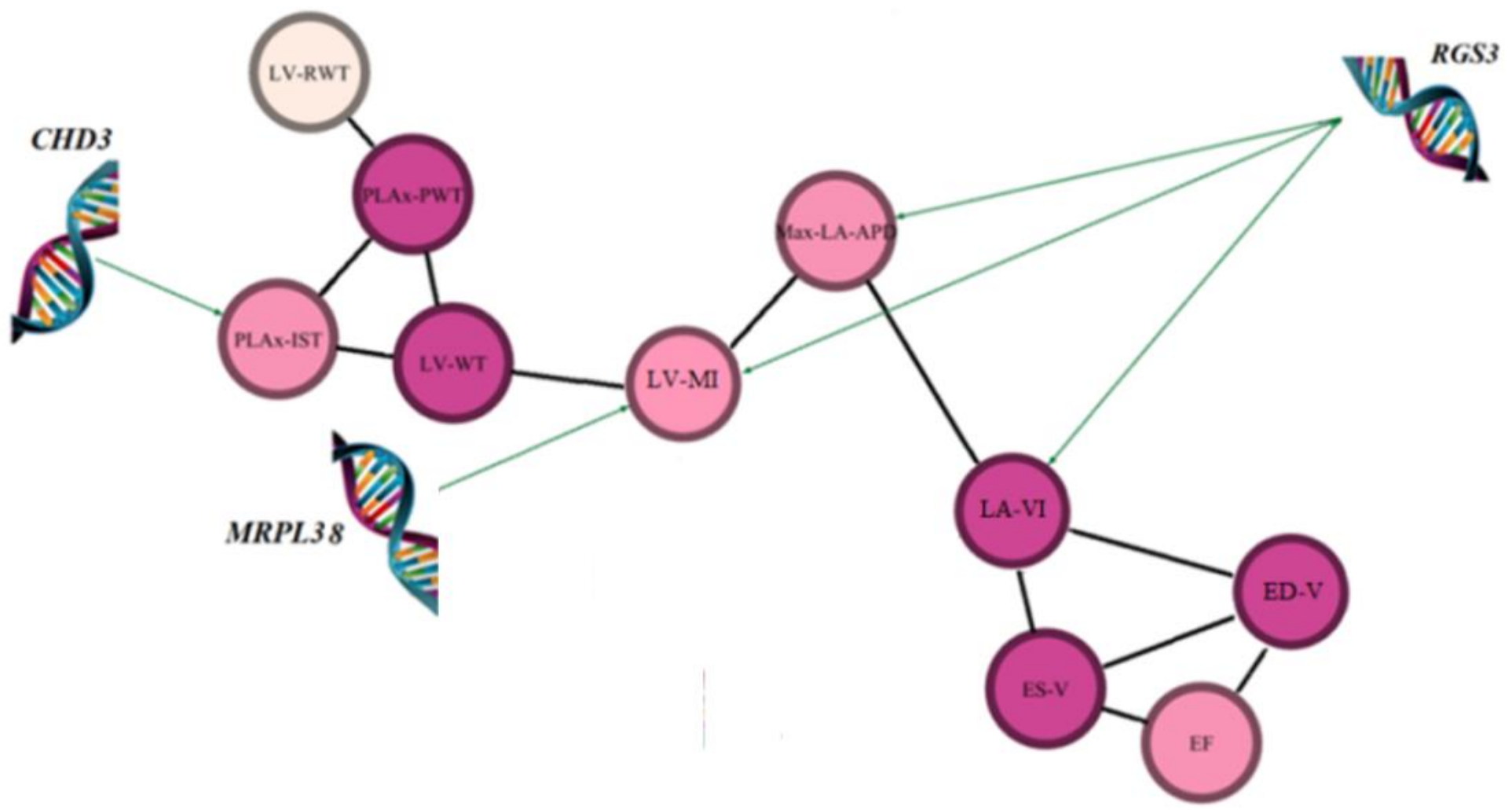

**Figure 4:** Identified genetic pathway to cardiac structure and function using IBMT.

**Table 2.** Selected genomic variants related to the traits, using a 98% Bayesian credible interval. HGVS name is description of sequence variation in genomic established by The Human Genome Variation Society; refSNP ID is a unique identifier provided by NCBI; CHR is the chromosome number; and MAF stands for minor allele frequency.

| **Identified variants** | | | | | |
|---|---|---|---|---|---|
| **HGVS name** | **refSNP ID** | **CHR** | **MAF%** | **Gene name** | **Related trait** |
| NC_000009.11:g.116346115C>A | rs144636307 | 9 | 0.38 | *RGS3* | LV-MI<br>LA-VI<br>Max-LA-APD |
| NC_000017.10:g.7802658C>T | rs200287864 | 17 | 0.25 | *CHD*3 | PLAx-IST |
| NC_000017.10:g.73897977C>T | rs76054219 | 17 | 0.32 | *MRPL38* | LV-MI |

Figure 5 shows empirical distributions of LV-MI (red/blue) for individuals with reference/alternative allele of one of the identified variants (NC_000017.10:g.73897977C>T). The noticeable shift of the distribution for individuals with mutation is observable. Supplementary Figure 2 shows levels of the traits for individuals with identified rare mutation over distribution of the traits.

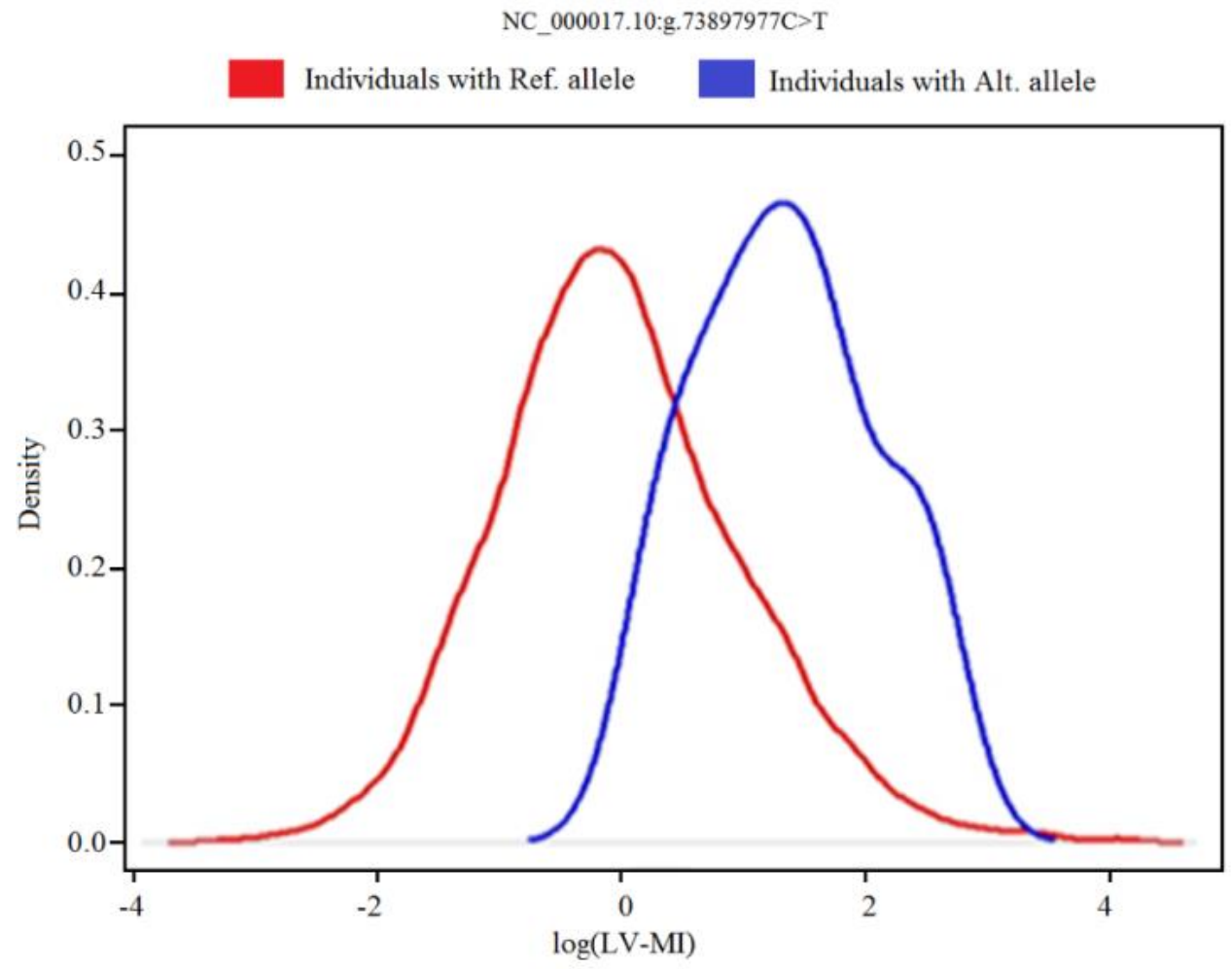

**Figure 5.** Empirical Distributions of log(LV-MI) for individuals with reference/alternate alleles for NC_000017.10:g.73897977C>T variant.

**Table 3.** Estimated effect (Est-Eff) and standard deviation (SD-Eff) of the identified genes with significant effect on the traits.

| HGVS name | Trait | Est-Eff | SD-Eff |
|---|---|---|---|
| NC_000009.11:g.116346115C>A | LV-MI | 1.13 | 0.104 |
| NC_000009.11:g.116346115C>A | LA-VI | 1.27 | 0.110 |
| NC_000009.11:g.116346115C>A | Max-LA-APD | 1.29 | 0.102 |
| NC_000017.10:g.7802658C>T | PLAx-IST | 1.48 | 0.186 |
| NC_000017.10:g.73897977C>T | LV-MI | 1.28 | 0.153 |

One of the identified variants located in chromosome 9 (NC_000009.11:g.116346115C>A) showed pleotropic effect on LV-MI, LA-VI and Max-LA-APD. Among 14 individuals (5 Females and 9 Males) having rare mutation in *RGS3* (NC_000009.11:g.116346115C>A) with pleiotropic action, two of the females and one male showed different patterns with other individuals, such that their level of LV-MI, LA-VI and Max-LA-APD are not greater than third quartiles (Supplementary Figure 2). This suggests that the rare mutation identified in *RGS3* may have higher impact on males. In addition, pleiotropic effect of this variant may represent different functions of the *RGS3* gene.

According to the Variant Effect Predictor (VEP) analysis (30), this variant is most likely a missense mutation in *RGS3* (regulator of G protein signaling 3) gene. It yields a codon change from ACC to AAC replacing the amino acid Threonine (Thr) by an Asparagine (Asn) in the protein sequence. *RGS* genes encode proteins that act as GTPase activating proteins (GAPs) and down-regulate G protein signaling. More details about the results from the VEP analysis is provided in Supplementary Table 1.

Another identified variant with impact on PLAx-IST is in chromosome 17 (NC_000017.10:g.7802658C>T). Individuals with T allele of this rare variant, including 2 Females and 8 Males, all have PLAx-IST level greater than third quartile of the trait (Supplementary Figure 2).

The variant NC_000017.10:g.7802658C>T is intronic to Chromatin Helicase DNA Binding Protein 3 (*CHD3*) gene based on VEP analysis (Supplementary Table 2). *CHD3* belongs to a family of genes coding for proteins that bear CHROMO (chromatin organization modifier) and SNF2-related helicase/ATPase domains. *CHD3* protein is one of the components of the Mi-2/NuRD (histone deacetylase) complex that participates in the remodeling of chromatin structure via histone deacetylation.

The third identified genetic variant influencing LV-MI is located in chromosome 17, gene *MRPL38* (NC_000017.10:g.73897977C>T). All 8 male individuals with rare mutation in *MRPL38* have LV-MI level greater than third quartiles of distribution (Supplementary Figure 2). However, the 2 female individuals do not show any different pattern.

The variant NC_000017.10:g.73897977C>T can be in a non-coding exon or be a missense mutation of *MRPL38* gene. The missense mutation of CGG into CAG codon causes the substitution of amino acid Arginine (Arg) by a Glutamine (Gln) concluded. Gln is a non-charged amino acid and smaller than Arg with putatively less capacity to create hydrogen bonds and favorable electrostatic interactions than Arg (see Supplementary Table 3 for VEP results). *MRPL38* encodes mitochondrial ribosomal proteins (*MRP*). Family *MRP*s stabilize mitochondrial ribosome (mitoribosome) and are responsible for the mitochondrial translation of 13 protein components of the Oxidative Phosphorylation (OXPHOS) gene complex in the mitochondrial DNA (31).

**Discussion**

Single trait analysis did not identify any genetic variants with significant impact on the 10 considered traits of cardiac structure and function in European American individuals from ARIC study. Therefore, we integrated Bayesian network and Bayesian multi-trait approach to improve the performance of the analysis. This Integrative Bayesian Multi-Trait (IBMT) approach provides sparse precision matrix and, eventually, more precise estimates of parameters in the model. Utilizing the IBMT method to increase the power of identification, in addition to carefully adjusting for covariates to avoid data contamination, and choosing the appropriate transformation function for each trait, we identified three significant genetic variants. These variants located in *RGS3*, *CHD3*, and *MRLP38* genes are rare which requires a high statistical power to detect any association with trait(s) (32)(33)(34)(35). Among those, the variant NC_000009.11:g.116346115C>A in *RGS3* showed pleiotropic action on vertical mass index, left vertical volume index, Maximum left atrial anterior-posterior diameter (Figure 4).

The variant NC_000009.11:g.116346115C>A is in the exon of gene *RGS3* (Regulator of G protein signaling 3) which belongs to *RGS* family. *RGS* family codes for proteins that act as GAPs and down-regulate G protein signaling. Many studies have proven that *RGS* gene expression is highly regulated in myocardium (36)(37)(38). Quantitative messenger RNA (mRNA) analysis revealed that *RGS3* is most highly expressed in human heart (39)(40)(41). The N-terminus of *RGS3* can inhibit TGFβ induced differentiation of pulmonary fibroblasts which is associated with left ventricular dilation and systolic dysfunction (42)(43)(44). The identified variant in *RGS3* was associated with pleotropic effect on LV mass (LV-MI) and measures of left atrial size (LA-VI and Max-LA-APD). The left atrial size may reflect the cumulative effects of increased LV filling pressure and diastolic function (45) and is a predictor of heart failure, ischemic stroke, and death. Thus, genetic variants contributing to abnormalities of LV mass and worsened diastolic function would be expected to potentially be associated with LA size.

As missense mutation, NC_000009.11:g.116346115C>A yields a codon change from ACC to AAC, which replaces a Threonine (Thr) to an Asparagine (Asn) in the protein amino acid sequence. The change from Thr to Asn does not alter side chain electrostatic charge because of both amino acids being electrostatically neutral. This could eventually affect hydrogen bond pattern since Asn has an extra hydrogen bond donor group (NH2). However, it is difficult to evaluate the final effect in the protein three dimensional structure since there may be alternative spliced transcripts. If this mutation happened at the interaction interface between *RGS3* and Gα subunit, it could eventually affect GTPase activity.

The variant NC_000017.10:g.7802658C>T that is significantly associated with parasternal long axis interventricular septum thickness (PLAx-IST) is in an intron of the *CHD3* gene, which codes for the Chromatin Helicase DNA Binding Protein 3 as a part of the chromatin structure remodeling complex. This variant is a potential splicing variant that could affect the rate of mature mRNA synthesis, and ultimately impact gene transcription.

The other identified variant NC_000017.10:g.73897977C>T is within *MRPL38* gene, a member of Mitochondrial ribosomal proteins (MRP) family that are part of the large subunit of the mitochondrial ribosome. As missense mutation, the change of CGG into CAG codon causes the substitution of amino acid Arginine (Arg) by a Glutamine (Gln). Gln is a non-charged amino acid and smaller than Arg with putatively less capacity to create hydrogen bonds and favorable electrostatic interactions than Arg. If the amino acid mutation takes place at the interface between *MRPL38* and mitochondrial ribosome, it could decrease binding affinity and destabilize mitochondrial ribosome. This in turn may reduce ribosomal protein synthesis levels and affect the oxidative phosphorylation pathway.

Overall, this study suggests that rare mutation might provide a better understanding of genetic impact on cardiovascular structure and resulting in remodeling cardiovascular disease and/or heart failure, although the analysis requires new approaches.

**Acknowledgements**
Thanks go to Dr. Boerwinkle for providing the data. Thanks also go to the staff and participants of the Atherosclerosis Risk in Communities (ARIC) Study for gathering the data. The ARIC Study is a collaborative study supported by the National Heart, Lung, and Blood Institute, National Institutes of Health, Contracts HHSN268201100005C, HHSN268201100006C and HHSN26-82011-00008C.

**Competing interests:** The author(s) declare no competing interests.

## Supplementary Figures

### Supplementary Figure 1

The histogram of the traits in Table1 after winsorization, standardization, and normal transformation

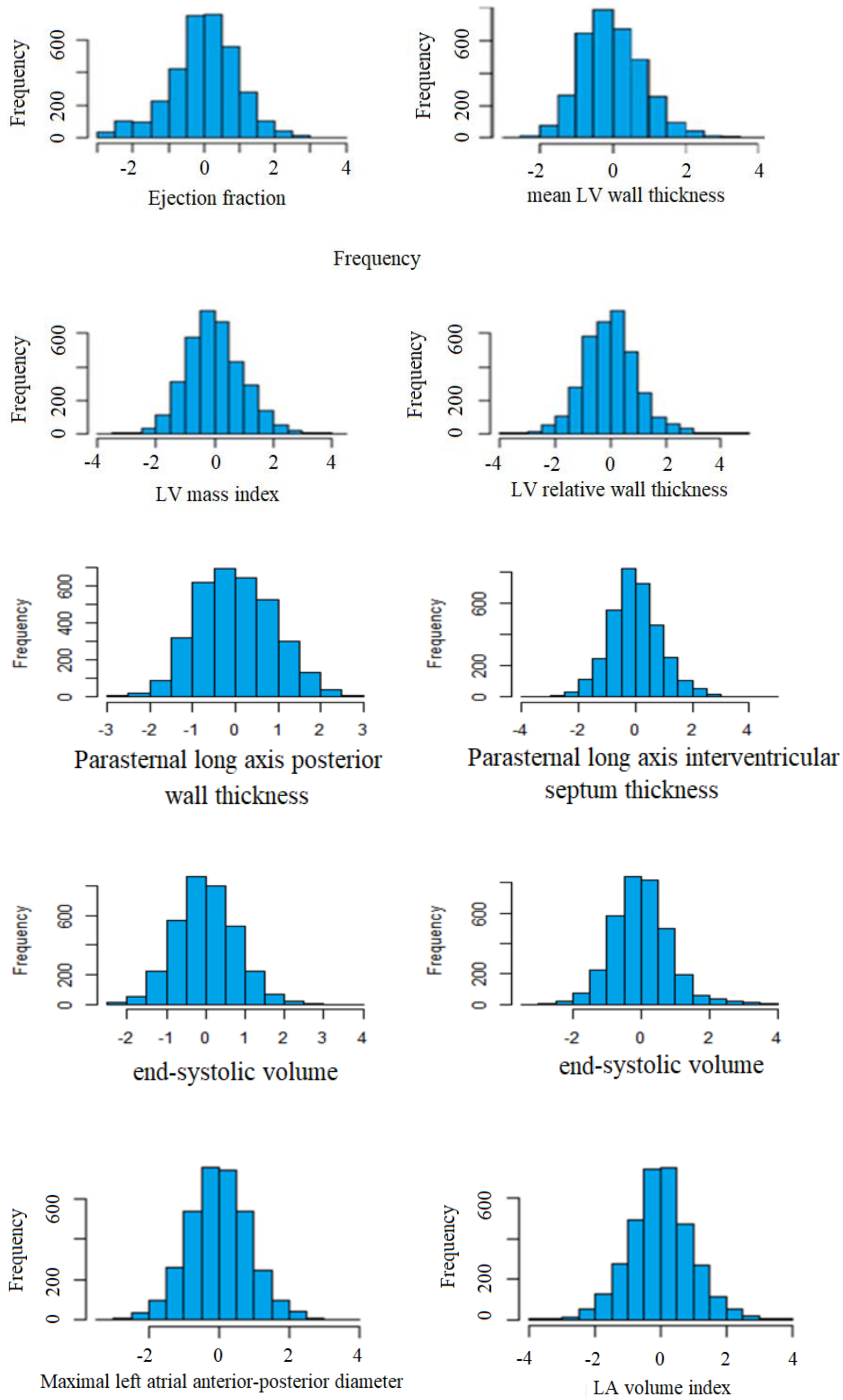

**Supplementary Figure 2**

Empirical Distributions of the traits in Table 3 for individuals with Reference allele: The yellow vertical line shows the third quartiles of the distribution. Individuals with alternative allele are represented with blue vertical line.

**NC_000009.11:g.116346115C>A in *RGS3***

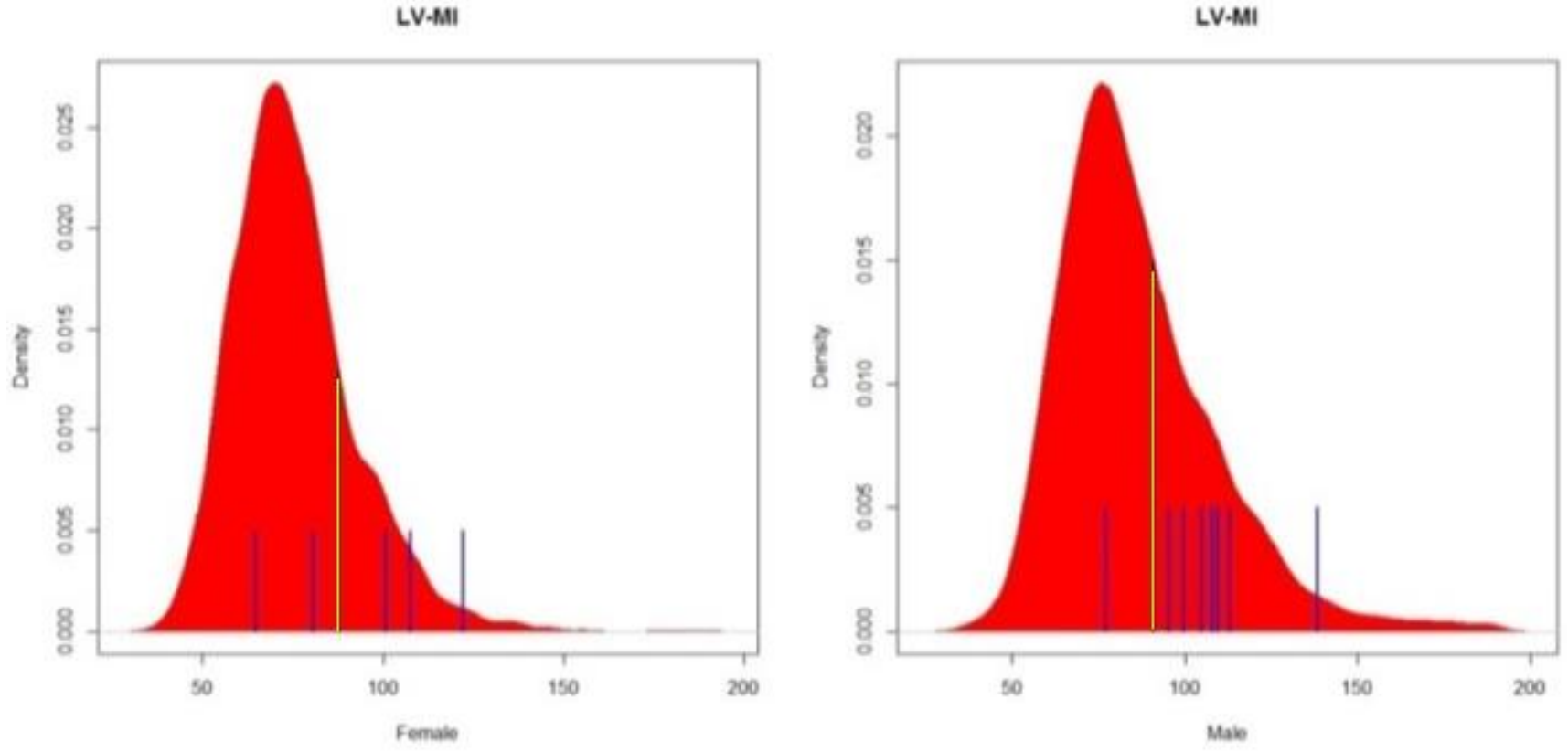

**NC_000009.11:g.116346115C>A in *RGS3***

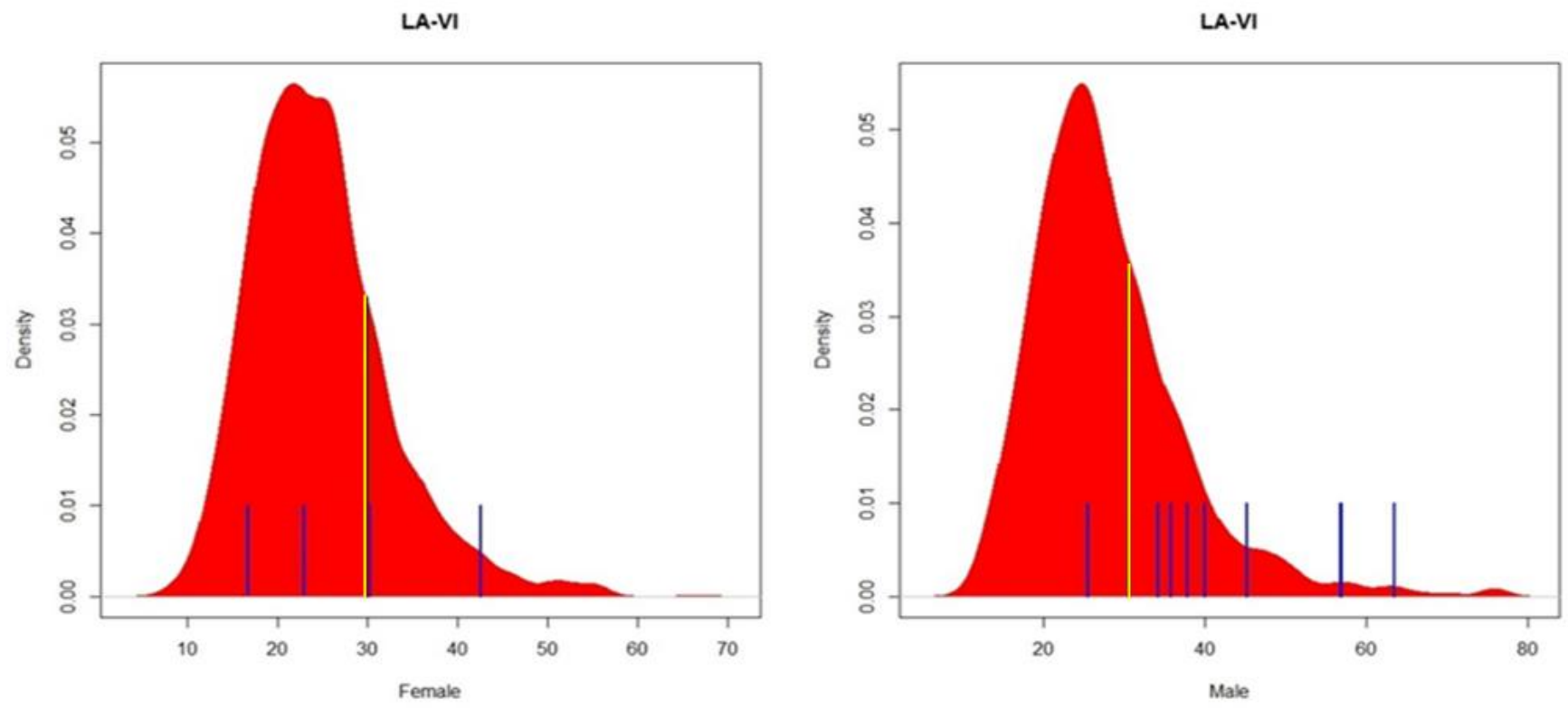

**NC_000009.11:g.116346115C>A in *RGS3***

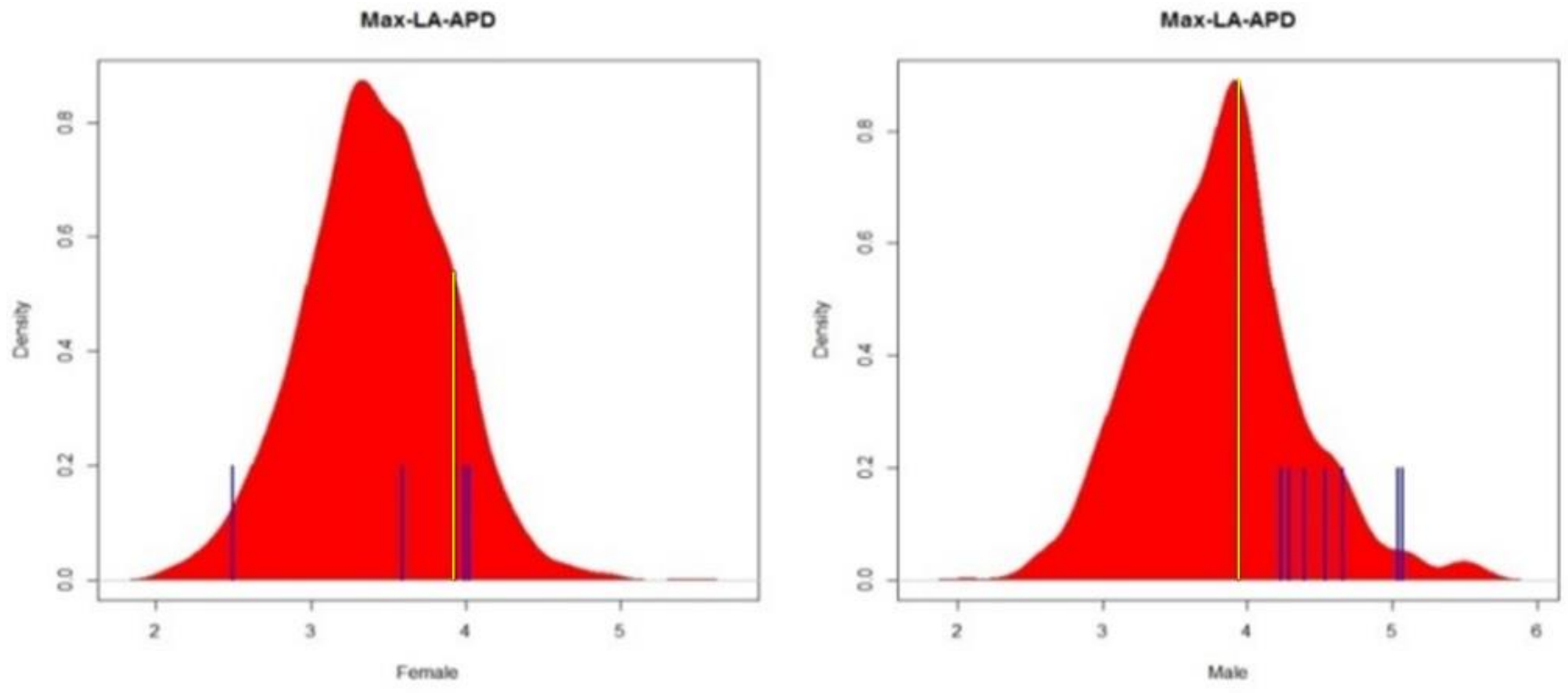

**NC_000017.10:g.7802658C>T in CHD3**

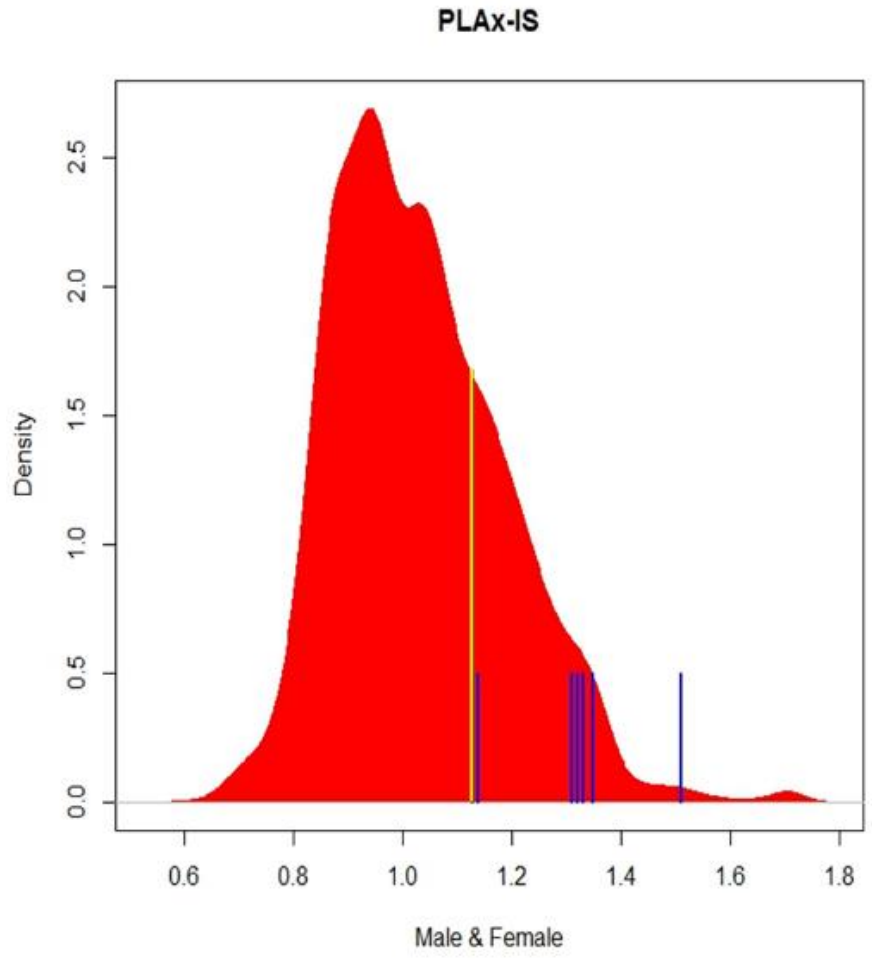

**NC_000017.10:g.73897977C>T in *MRPL38***

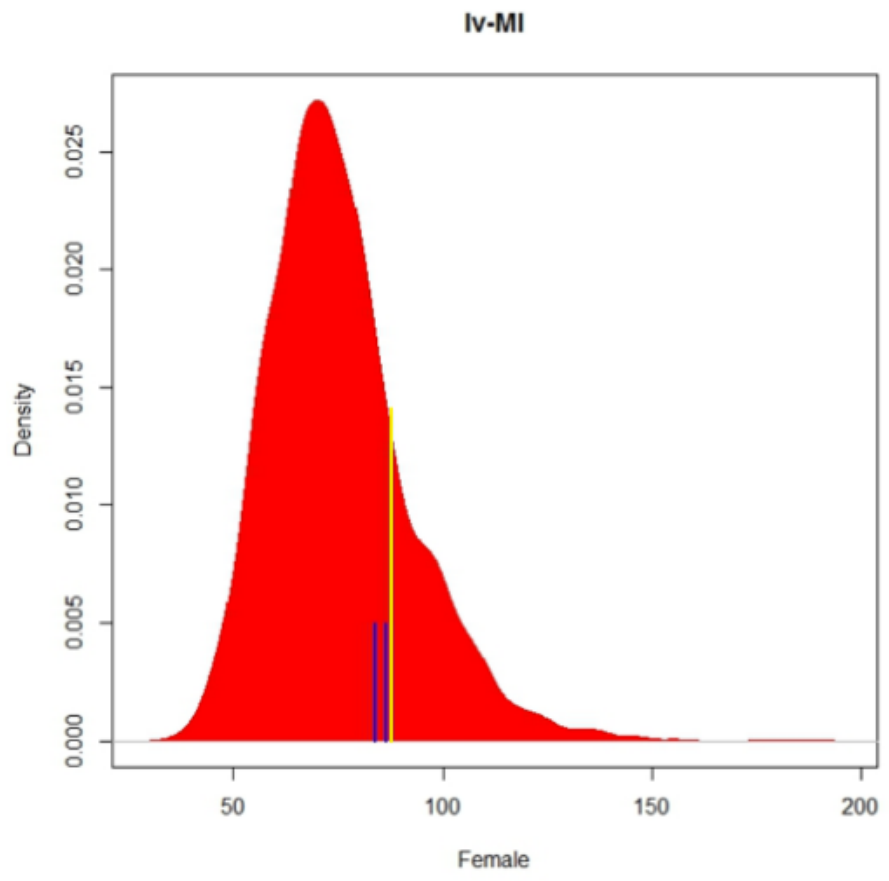

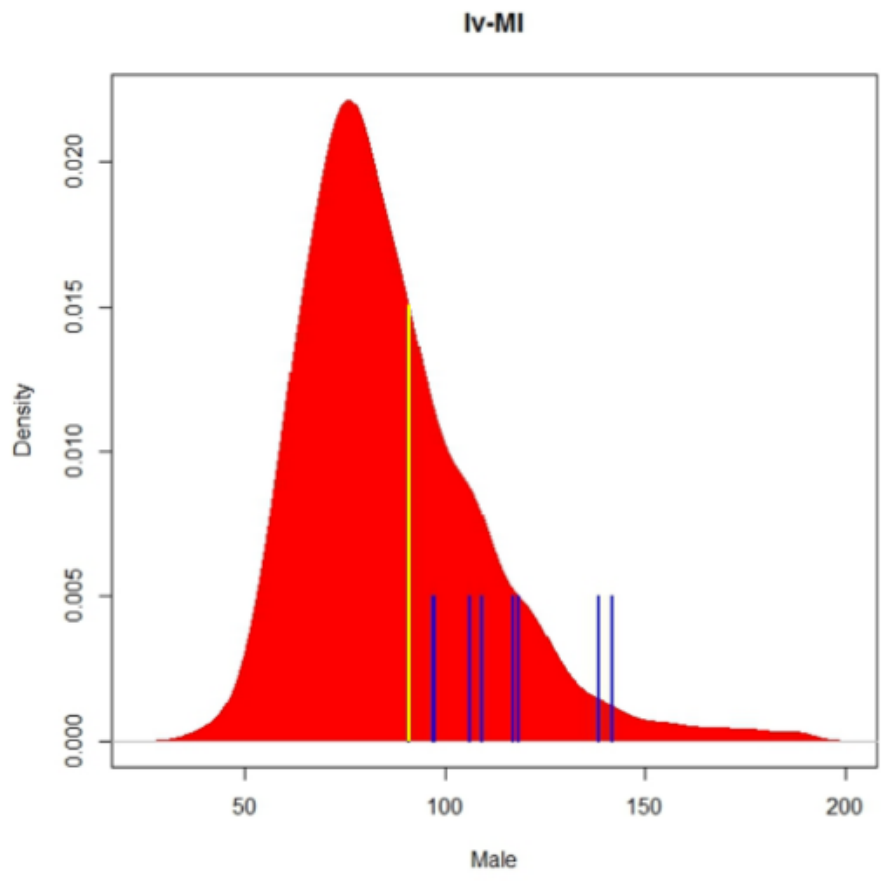

## Supplementary Tables

The VEP analysis results for identified genetic variants are summarized in the following tables where

- **HGVS name**: description of sequence variation in genomic established by the Human Genome Variation Society
- **Consequence:** consequence type of the allele on transcript
- **AA_AF/EA_AF:** African Americans/ European Americans allele frequency in NHLBI-ESP data base
- **HGVSp:** HGVS protein sequence name

## Supplementary Table 1

| HGVS name | Allele | Consequence | Gene symbol | Biotype | Exon | Intron | HGVSp | Amino acids | Codons |
|---|---|---|---|---|---|---|---|---|---|
| NC_000009.11: g.116346115C>A | A | missense_variant | RGS3 | protein_ coding | 2/7 | - | NP_001263189.1: p.Thr129Asn | T/N | aCc/aAc |
| NC_000009.11: g.116346115C>A | A | intron_variant | RGS3 | protein_ coding | - | 10/14 | - | - | - |
| NC_000009.11: g.116346115C>A | A | missense_variant | RGS3 | protein_ coding | 3/8 | - | NP_001269851.1: p.Thr129Asn | T/N | aCc/aAc |
| NC_000009.11:g. 116346115C>A | A | missense_variant | RGS3 | protein_ coding | 18/23 | - | NP_001269852.1: p.Thr698Asn | T/N | aCc/aAc |
| NC_000009.11: g.116346115C>A | A | missense_variant | RGS3 | protein_ coding | 2/7 | - | NP_001309144.1: p.Thr147Asn | T/N | aCc/aAc |
| NC_000009.11: g.116346115C>A | A | missense_variant | RGS3 | protein_ coding | 11/16 | - | NP_570613.2: p.Thr527Asn | T/N | aCc/aAc |
| NC_000009.11: g.116346115C>A | A | intron_variant | RGS3 | protein_ coding | - | 1/4 | - | - | - |
| NC_000009.11: g.116346115C>A | A | missense_variant | RGS3 | protein_ coding | 21/26 | - | NP_652759.3: p.Thr808Asn | T/N | aCc/aAc |
| NC_000009.11: g.116346115C>A | A | non_coding_ transcript_ exon_variant | RGS3 | misc_RNA | 18/23 | - | - | - | - |
| NC_000009.11: g.116346115C>A | A | non_coding_ transcript_ exon_variant | RGS3 | misc_RNA | 3/8 | - | - | - | - |

## Supplementary Table 2

| HGVS name | Allele | Consequence | Gene symbol | Biotype | Exon | Intron | HGVSp | Amino acids | Codons |
|---|---|---|---|---|---|---|---|---|---|
| NC_000017.10: g.7802658C>T | T | splice_region_ variant,intron_ variant | CHD3 | protein_ coding | - | 14/39 | - | - | - |
| NC_000017.10: g.7802658C>T | T | splice_region_ variant,intron_ variant | CHD3 | protein_ coding | - | 14/39 | - | - | - |
| NC_000017.10: g.7802658C>T | T | splice_region_ variant,intron_ variant | CHD3 | protein_ coding | - | 14/38 | - | - | - |
| NC_000017.10: g.7802658C>T | T | splice_region_ variant,intron_ variant | CHD3 | protein_ coding | - | 14/39 | - | - | - |
| NC_000017.10: g.7802658C>T | T | splice_region_ variant,intron_ variant | CHD3 | protein_ coding | - | 14/38 | - | - | - |
| NC_000017.10: g.7802658C>T | T | splice_region_ variant,intron_ variant | CHD3 | protein_ coding | - | 14/39 | - | - | - |
| NC_000017.10: g.7802658C>T | T | splice_region_ variant,intron_ variant | CHD3 | protein_ coding | - | 14/33 | - | - | - |
| NC_000017.10: g.7802658C>T | T | splice_region_ variant,intron_ variant | CHD3 | protein_ coding | - | 14/21 | - | - | - |

## Supplementary Table 3

| HGVS name | Allele | Consequence | Gene symbol | Biotype | Exon | Intron | HGVSp | Amino acids | Codons |
|---|---|---|---|---|---|---|---|---|---|
| NC_000017.10: g.73897977C>T | T | upstream_gene_ variant | TRIM65 | protein_ coding | - | - | - | - | - |
| NC_000017.10: g.73897977C>T | T | missense_variant | MRPL38 | protein_ coding | 4/9 | - | - | R/Q | cGg/cAg |
| NC_000017.10: g.73897977C>T | T | upstream_gene_ variant | TRIM65 | protein_ coding | - | - | - | - | - |
| NC_000017.10: g.73897977C>T | T | non_coding_ transcript_ exon_variant | MRPL38 | misc_RNA | 4/8 | - | - | - | - |

**Supplementary statistical methods**

Let assume that we measured q phenotypic traits for n individuals with p recorded genotypic variants. The multivariate polygenic model for this data can be represented as

$$\boldsymbol{Y}_{\mathrm{nq}\times 1} = \boldsymbol{\mu}_{\mathrm{nq}\times 1} + \left(\boldsymbol{X}_{\mathrm{n}\times \mathrm{p}} \otimes \boldsymbol{I}_{\mathrm{q}}\right)\boldsymbol{\beta}_{\mathrm{pq}\times 1} + \boldsymbol{U}_{\mathrm{nq}\times 1} + \boldsymbol{\epsilon}_{\mathrm{qn}\times 1} \qquad (1)$$

where

$$\boldsymbol{\epsilon} \sim N(\mathbf{0}, \Sigma),\ \Sigma_{nq\times nq} = \mathrm{diag}[\Sigma_{ii}]_{q\times q},\ \Sigma_{ii} = \mathrm{diag}[\sigma_i]_{n\times n}.$$

Each entries of $\boldsymbol{Y} = \left\{y_{ij}\right\}_{\substack{i=1,\dots,q \\ j=1,\dots,n}}$ represents ith trait recorded for jth individual, $\beta_{ik}$ is the effect of kth genomic variants on ith trait in coefficient vector $\boldsymbol{\beta} = \{\beta_{ik}\}_{\substack{i=1,\dots,q \\ k=1,\dots,p}}$ , vector $\boldsymbol{U} = \left\{u_{ij}\right\}_{\substack{i=1,\dots,q \\ j=1,\dots,n}}$ includes random effects corresponding to vector Y as

$$\boldsymbol{U} \sim N(\mathbf{0}, \Psi)$$

where

$$\Psi = \begin{bmatrix} \Psi_{11} & \cdots & \Psi_{1q} \\ \vdots & \ddots & \vdots \\ \Psi_{q1} & \cdots & \Psi_{qq} \end{bmatrix},$$

and $\Psi_{il} = \left\{\psi_{jg}\right\}_{\substack{j=1,\dots n, \\ g=1,\dots,n}}, \quad i, l \in \{1, \dots, q\}$.

Without loss of generality, we assume $y_{ij}$s are standardized and write the likelihood function as

$$\ell(\boldsymbol{\beta}, \boldsymbol{U}, \Sigma, \Psi\ |\boldsymbol{Y}) \propto \exp\left\{-\frac{1}{2}\ \boldsymbol{e}^T \Sigma^{-1} \boldsymbol{e}\right\}$$

where $\boldsymbol{e} = \boldsymbol{Y} - \left(\boldsymbol{X} \otimes \boldsymbol{I}_q\right)\boldsymbol{\beta} - \boldsymbol{U}$.

**Bayesian network over traits**: The large sample size (n) and large number of traits (q) introduce a large number of parameters into the model through $\Psi$ and make the numerical algorithm of model (1) unstable and slow to converge. Therefore, to reduce the number of parameters in polygenic mixed model and provide larger precision for parameter estimations, we propose to estimate the sparse structure of precision matrix $\Psi^{-1}$ in priori using phenotypic data (individuals without genotype record) which is accessible in majority of medical studies. Here in particular, we apply a Bayesian network that relies on probabilistic graphical models to estimate sparse relationship among traits.

Bayesian networks represent underlying relationship among traits based on efficient and effective representation of their joint probability distribution (1)(2)(3) where nodes represent traits and links represent partial correlations. A missing edge between two traits reveals that the two corresponding traits do not have a significant relationship after excluding effect of the other traits in the analysis. To learn about sparsity of precision matrix $\Psi^{-1}$, we incorporate independencies inferred from Bayesian networks to reduce the number of parameters in the model and efficiently compute posterior probabilities. In the matrix $\Psi^{-1}$, we replace zero with the parameters corresponding to missing arrows in the identified Bayesian network over the traits. To avoid overfitting, we identify the Bayesian network using the data on the set of individuals without genotype record.

**Prior Specification:** There have been many studies on setting a prior distribution on precision matrix $\Psi^{-1}$. The most common approach is to set Wishart distribution as prior on $\Psi^{-1}$ while it is conjugate prior for multivariate normal model. Wishart distribution is fully parameterized with a single degree of freedom parameter and scale matrix parameter. The degree of freedom parameter that needs to be larger than dimension of underling covariance matrix represents the strength of information surrounded around scale matrix parameter. Therefore, in large scale problems, this choice of prior yields to highly concentrated distribution about the scale matrix(4)(5)(6).

The assumption of independent individuals that is held in many genetic studies leads to a sparse covariance matrix $\Psi$ as

$$\Psi = \begin{bmatrix} \Psi_{11} & \cdots & \Psi_{1q} \\ \vdots & \ddots & \vdots \\ \Psi_{q1} & \cdots & \Psi_{qq} \end{bmatrix}, \qquad \Psi_{il} = \mathrm{diag}\left[\psi_{jg}\right], \;\; i,l \in \{1,\dots,n\}, \;\;\; j,g \in \{1,\dots,q\}.$$

To set prior on this matrix, we first rearrange vector U in order to be partition with individual's index j. If we denote the rearranged random effect vector with $\boldsymbol{U}^*$ such that $\boldsymbol{U}^* \sim \mathrm{N}(\mathbf{0}, \Psi^*), \Psi^*$ is a block diagonal as

$$\Psi^* = \begin{bmatrix} \Psi_{11}^* & \cdots & 0 \\ \vdots & \ddots & \vdots \\ 0 & \cdots & \Psi_{nn}^* \end{bmatrix}$$

where each $\Psi_{jj}^*$ is a $q \times q$ dense matrix. $\Psi_{jj}^*$ takes into account correlation among different traits for jth individual. While $\Psi^{*-1}$ is a block diagonal matrix and q is a small number, the use of Wishart prior for $\Psi_{jj}^{*-1}$is appropriate. Furthermore, we incorporate the identified relationship among the traits using Bayesian network and estimate the sparsity of $\Psi_{jj}^{*-1}$and set G-Wishart (GW) distribution (7) as prior on $\Psi_{jj}^{*-1}$. G-Whishart distribution, which is a restricted domain of Wishart given constraints of network structure, provides sparse block diagonal matrix and reduces computational burden time. The density of G-Wishart is

$$p\left(\Psi_{jj}^{*-1} | \mathrm{BN_T}\right) = \left(I_{G_T}(\nu, \Lambda)\right)^{-1} \left|\Psi^{*-1}{}_{jj}\right|^{\frac{\nu-2}{2}} \exp\left\{-\frac{1}{2}\mathrm{tr}\left(\Lambda \Psi^{*-1}{}_{jj}\right)\right\}$$

where $\mathrm{BN_T}$ stands for the Bayesian Networks over the traits, and

$$I_{GT}(\nu, \Lambda) = \int |\Psi_{jj}^{*-1}|^{(\nu-2)/2} \; \exp\left\{-\frac{1}{2}\mathrm{tr}\left(\Lambda \Psi_{jj}^{*-1}\right)\right\} \, \mathrm{d}\Psi_{jj}^{*-1}$$

is the normalizing constant, which is finite for $\nu > 2$ .This results in a reduction of the number of parameters in polygenic mixed model, and consequently expedites the convergence of the algorithm.

We set conjugate prior for other hyperparameters of the model as

$$\boldsymbol{\beta} \sim \mathrm{N}\ (0, \boldsymbol{\Omega})$$

where

$$\Omega_{pq\times pq} = \text{diag}\ [\Omega_{kk}]_{p\times p}, \qquad \Omega_{kk} = \text{diag}\ [\sigma_j]_{p\times p}$$

and

$$\sigma_i \sim IG\ \left(\frac{a_1}{2}, \frac{b_1}{2}\right)$$

The aforementioned prior specification leads to Gibbs sampling scheme for the model with the graphical representation shown in the following Figure.

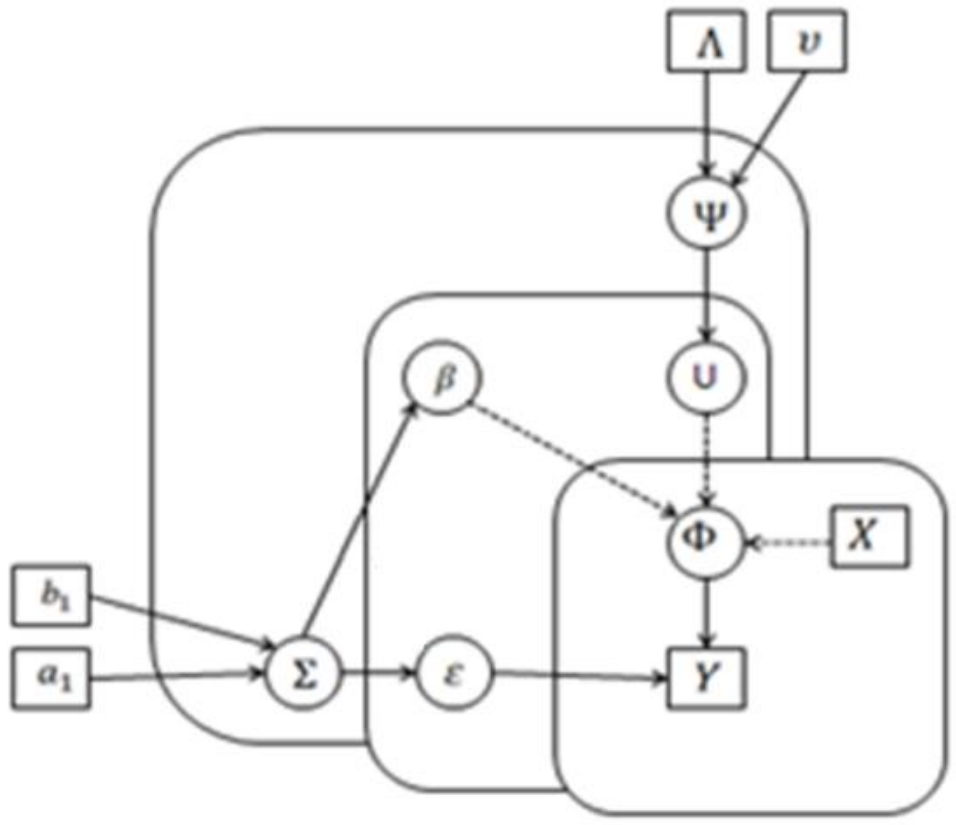

Graphical representation of the model

**Gibbs Sampling Scheme of IBMT:**

$\boldsymbol{\beta}_i \mid \boldsymbol{y}_i, \boldsymbol{u}_i, \boldsymbol{\sigma}_i \sim \mathrm{N}\ [\ (\boldsymbol{X}^T\boldsymbol{X} + \boldsymbol{I})^{-1}\boldsymbol{X}^{-1}(\boldsymbol{y}_i - \boldsymbol{u}_i), \sigma_i(\boldsymbol{X}^T\boldsymbol{X} + \boldsymbol{I})^{-1}]$

$\mathbf{u}_{\mathrm{j}}^* \mid \mathbf{u}_{\mathrm{j}}, \boldsymbol{\Psi}_{\mathrm{jj}}^*, \boldsymbol{\Sigma}_{\mathrm{jj}} \sim \mathrm{N}\ \left[\left(\boldsymbol{\Psi}_{\mathrm{jj}}^* + \boldsymbol{\Sigma}_{\mathrm{jj}}^{-1}\right)^{-1} \boldsymbol{\Sigma}_{jj}^{-1}\mathbf{z}_j, \qquad \left(\Psi_{\mathrm{jj}}^* + \Sigma_{jj}^{-1}\right)^{-1}\right]$

$\sigma_i \mid \boldsymbol{y}_i, \boldsymbol{\beta}_i, \boldsymbol{u}_i \sim \mathrm{ING}\left[\frac{n+p+a_1}{2}, \frac{1}{2}(\boldsymbol{e}_i{}^T\boldsymbol{e}_i + \boldsymbol{\beta}_i^T\boldsymbol{\beta}_i + b_1)\right]$

$\boldsymbol{\Psi}_{jj}^{*-1} \mid \boldsymbol{u}_j \sim G\mathrm{W}[\boldsymbol{\Lambda} + \boldsymbol{u}_j^T\boldsymbol{u}_j, \qquad \nu + q]$

where $\boldsymbol{e_i} = (\boldsymbol{Y_i} - \boldsymbol{X\beta_i} - \boldsymbol{U_i})$ and vector Z is obtained after reordering entries of vector $(\boldsymbol{Y} - \boldsymbol{X} \otimes \boldsymbol{I}_q)\ \boldsymbol{\beta}$ in order to be partitioned by each individual similar to $\boldsymbol{U}^*$.

**Supplementary references**